# Thermal transport in silicene nanotubes: Effects of length, grain boundary and strain

Maryam Khalkhali[1], Farhad Khoeini[1]*, Ali Rajabpour[2]

[1.] Department of Physics, University of Zanjan, Zanjan 45195-313, Iran

[2.] Mechanical Engineering Department, Imam Khomeini International University, Qazvin 34148–96818, Iran

**Abstract**
Thermal transport behavior in silicene nanotubes has become more important due to the application of these promising nanostructures in the engineering of next-generation nanoelectronic devices. We apply non-equilibrium molecular dynamics (NEMD) simulations to study the thermal conductivity of silicene nanotubes with different lengths and diameters. We further explore the effects of grain boundary, strain, vacancy defect, and temperature in the range of 300-700 K on the thermal conductivity. Our results indicate that the thermal conductivity varies with the length approximately in the range of 24-34 W/m.K but exhibits insensitivity to the diameter and chirality. Besides, silicene nanotubes consisting of the grain boundary exhibit nearly 30% lower thermal conductivity compared with pristine ones. We discuss the underlying mechanism for the conductivity suppression of the system consisting of the grain boundary by calculating the phonon power spectral density. We find that by increasing the defect concentration and temperature, the thermal conductivity of the system decreases desirably. Moreover, for strained nanotubes, we observe unexpected changes in the thermal conductivity, so that the conductivity first increases significantly with tensile strain and then starts to decrease. The maximum thermal conductivity for the armchair and zigzag edge tubes appears at the strains about 3% and 5%, respectively, which is about 28% more than that of the unstrained structure.

**Introduction**
Nowadays, silicene has been the focus of extensive research due to its remarkable and exotic properties, such as possessing the flexible crystal structure [1], the large spin-orbit coupling [2], electrically tunable band gap [3], half-metallicity [4], quantum Hall effect [5], and superconductivity [6]. The theoretical possibility of the existence of silicene sheet was determined with first-principles calculations in 1994 [7]. Despite the theoretical improvements, experimental studies on silicene did not exist until 2010, when silicene was fabricated by epitaxial growth on a silver surface [8, 9].
The compatibility and adaptability of silicene with silicon-based nanotechnology make it suitable for next-generation nanoelectronic devices such as Li-ion batteries [10, 11], reusable molecule sensors [12], field effect transistor [13], and high-performance spintronic devices [14].
Also, as regards of thermoelectric application, silicene is even more promising than graphene [15] because graphene exhibits ultra-high thermal conductivities of 2000-5300 W/m.K [16, 17].

However, due to the buckled structure of silicene, the symmetry selection rule that is used for graphene cannot be applied for silicene. Consequently, the out-of-plane modes experience more scattering channels, their contribution to the thermal conductivity is smaller than that of graphene, and overall thermal conductivity of silicene is impressively lower compared to that of the graphene [18]. Moreover, in the case of graphene, due to its zero energy gap, the Seebeck coefficient of graphene is very small [15]. In contrast, silicene with slightly buckled structure leads to a nonzero band gap that can be an efficient thermoelectric material for power generators or refrigeration applications [19].

Due to hardness's in the synthesis of free-standing silicene, there has not been any experimental achievement on the measuring the thermal conductivity of this material so far, but there are many works devoted to the numerical study of the thermal conductivity of silicene and reported magnitudes in the range of 5 to 69 W/m.K [18]. For instance, Hu et al. performed non-equilibrium molecular dynamics simulations to investigate thermal transport in a single-layer silicene sheet under uniaxial strain. The values of the thermal conductivity related to the unstrained systems were obtained to be 40.1 W/m.K for zigzag single-layer silicene, which was too lower than its counterpart graphene (2002.2 W/m.K) [20]. Pei et al. reported that the thermal conductivity of bulk silicene (41W/m.K) is much lower than that of bulk silicon (210W/m.K) [21].

Beside intense study on the features of silicene sheet and silicene nanoribbons, there have been many efforts directed at studies of silicene nanotubes, probing their unique properties and potential applications in nanotechnology [22-25]. As a pioneer study, Sha et al. reported the successful preparation of silicon nanotubes by the use of the chemical vapor deposition method by applying $Al_2O_3$ nanochannel in 2002 [22]. Recently, much interest is generated to develop the successful experimental process for synthesis the silicon nanotubes such as dc-arc plasma [26], dual RF plasma [27], hydrothermal [28], and template combined molecular beam epitaxy methods [29]. Epur et al. reported a new and straightforward approach including magnesium oxide nanorods for generating the hollow silicon nanotubes [30].

Also, intensive theoretical studies based on density functional theory and molecular dynamics simulation have been conducted to investigate the possibility of the existence of silicene nanotubes [31, 32] as well as the stability and geometric structure of single and double-walled silicone nanotubes [33]. Wang et al. studied the stability and electronic features of silicone nanotubes. They reported that the stability of chiral single and double-walled silicone nanotubes enhanced with increasing the diameter [34]. Liu et al. explored the electron properties of silicon nanotubes by density functional theory [35]. Salimian et al. applied the transfer matrix method to investigate the transfer properties of silicene nanotube field effect transistor. They reported that although carbon nanotubes are not appropriate for transistors action, silicene nanotubes, based on rolled-up silicene sheets, are applicable material because of having intrinsic band gap and silicene nanotube field effect transistor can switch between ON and OFF states [36].

Unlike intense studies on the mechanical and electronic feature of silicon nanotubes [37-40], limited research has been devoted to the thermal properties of silicene nanotubes until now. Therefore, because of the broad potential application of silicene-based nano-devices, studying the thermal transport behavior of silicene nanotubes and engineering its thermal conduction properties to use for a diverse range of applications may be an exciting challenge. Among various factors, the effect of strain, isotope doping, grain boundary, defects and impurity, crystal structures, temperature, torsion, and size on the thermal conductivity of nanostructures have attracted great deals of attention during the last decade [41-43].

The fact that strain can tune the thermal conductance of the silicene sheet and nanoribbons has been a topic for much research. Xie et al. predicted thermal conductivity of monolayer silicene under uniform bi-axial strain by performing first-principles calculations [44]. Hu et al. found unexpectedly change on the thermal conductivity of silicene with applying tensile strain [20].

In practice, nanosheets and nanotubes are mostly grown with different types of defects, which grain boundary is one of the most often types. Grain boundaries can affect the transport properties of nanostructures such as thermal conductivity. Bazrafshan et al. investigated the effect of grain size on the thermal conductivity of polycrystalline graphene by a multiscale method combined with non-equilibrium molecular dynamics (EMD) simulations and solving continuum heat conduction equation [45]. Cao et al. utilized MD simulations to investigate the effect of asymmetric tilt grain boundary between the armchair and zigzag graphene on the thermal conductivity of the sheet. They have found considerable temperature drop and thermal boundary resistance across the grain boundary [46].

In this study, non-equilibrium molecular dynamics simulations were carried out to calculate the thermal conductivity of zigzag and armchair silicene nanotubes with different lengths and diameters. Besides, the effects of the grain boundary between the armchair and zigzag oriented silicene nanotubes, strain, vacancy defect, and temperature (in the range of 300-700 K) on the thermal conductivity are explored. Furthermore, the density of states analysis is performed to investigate the influence of the grain boundary on the vibrational behaviors of atoms.

**Model and simulation method**
In semimetals, such as silicene, the contribution of electrons on the thermal conductivity at room temperature and above is less than that of phonons. Therefore, non-equilibrium molecular dynamics (NEMD) simulation that ignores electron transport can be used to investigate the thermal properties of silicene [47]. In this research, the non-equilibrium molecular dynamics (NEMD) simulation is carried out using Large-scale Atomic/Molecular Massively Parallel Simulator (LAMMPS) [48] to study the effect of the grain boundary, axial strain, random vacancy defect, and temperature on the thermal conductivity of silicene nanotubes. For this purpose, three different types of single-wall silicene nanotubes consisting of armchair silicene nanotube (ASNT), zigzag silicene nanotube (ZSNT), and silicene nanotube with Armchair-Zigzag grain boundary (A-Z GSNT) with different lengths and diameters are constructed and placed within a cubic simulation box.

For a precise explanation of the thermal properties of silicene nanotubes, choosing an appropriate potential function is essential. Previous studies [47, 49, 50] have shown that utilizing the Tersoff potential can better predict the equilibrium bond length between Si-Si atoms in silicene structure and the calculated feathers are in good agreement with the findings of first-principles calculations. Thus, we employed the Tersoff potential [51] to describe the Si-Si bonding interaction in silicene nanotubes. Newton's equations of motion are integrated via the velocity Verlet algorithm [52] with a time step of 1 fs, and the free boundary condition is employed in all directions.

First, the whole system is relaxed for 1 ns in NVT ensemble coupling to Nose-Hover temperature thermostat [53] at 300 K. Then, to generate temperature gradient and nonequilibrium heat current across the system, the silicene nanotube is divided into the N slabs along its length direction. Two outer slabs are fixed to prevent rotations of the system during the simulation time. Next to these fix regions, there are hot and cold slabs that are coupled to Nose-Hoover thermostat employing NVT ensemble for 3 ns at T+ΔT/2 (310 K) and T-ΔT/2 (290 K), respectively; where T is the mean temperature of the nanotube and ΔT is the temperature difference between the hot and the cold segments of the system, respectively. Furthermore, NVE ensemble is applied to integrate the equation of motion of atoms in all other slabs. After the system finds the steady state condition, the accumulative energy added into and subtracted from hot and cold regions were recorded every 1000 time steps and plotted versus time. The linear slopes of energy diagrams are equal to the heat current ($q_z$) of the system.

The thermal conductivity is calculated from one-dimensional Fourier's law in the z-direction as follows:

$$q_z = -KA\frac{\partial T}{\partial z}, \tag{1}$$

where $\frac{\partial T}{\partial z}$ is the effective temperature gradient along the z axis, $q_z$ is the calculated heat current, and $A$ is cross-section area of the silicene nanotube; the thickness of the silicene nanotube is supposed to be 4.65 Å since the van der Waals diameter of silicon atoms is 4.2 Å and buckling height of silicene is 0.45 Å.

All systems were simulated for the entire 3 ns after relaxation, where the first 1.5 ns was discarded as the pre-equilibration step. Also, to lower the statistical uncertainties of calculated results, each simulation was repeated with three different initial conditions. The schematic atomic configuration of A-Z GSNT and NEMD setup for calculation the thermal conductivity is shown in Fig. 1(a). In addition, views of ASNT and ZSNT perpendicular to the axial direction are represented in Figs. 1(b-c).

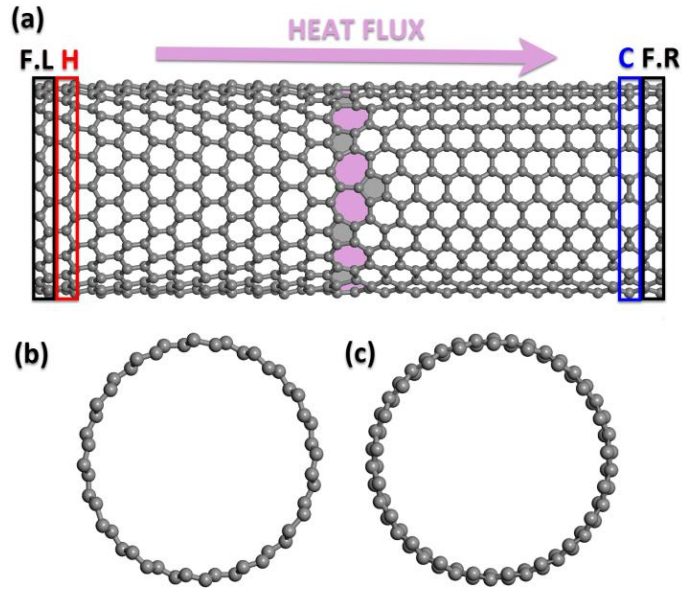

Fig. 1. (a) A schematic view of A-Z GSNT; (b-c) views of ASNT and ZSNT perpendicular to the axial direction

For the system with the grain boundary, when the heat flux is applied, a temperature drop $\Delta T_{in}$ is observed across the armchair-zigzag interface that causes the appearance of interface thermal resistance (Kapitza resistance) as [54]:

$$R_{in} = \frac{\Delta T_{in}}{J}, \tag{2}$$

where $J$ is heat flux.

In order to calculate the effective thermal conductivity of the system with armchair-zigzag grain boundary we have applied following mixture rule-based expression:

$$\kappa_{eff}^{-1} = \sum_i f_i \, \kappa_i^{-1}, \qquad (3)$$

where $f_i$ is volume fraction and $\kappa_i$ is the thermal conductivity of the ith part of the system [55]. According to this expression, the effective thermal conductivity of A-Z GSNT can be calculated by the following equation:

$$\kappa_{eff}^{-1} = \tfrac{1}{2}\kappa_{zigzag}^{-1} + \tfrac{1}{L}(G^{-1}) + \tfrac{1}{2}\kappa_{armchair}^{-1}, \qquad (4)$$

where $\kappa_{zigzag}$ and $\kappa_{armchair}$ are the thermal conductivities of the zigzag and armchair parts of A-Z GSNT, respectively, that can be calculated from Fourier's law. The temperature gradient ($\frac{\partial T}{\partial Z}$) in zigzag and armchair parts is approximated by $\frac{\Delta T_{zigzag}}{L_{zigzag}}$ and $\frac{\Delta T_{armchair}}{L_{armchair}}$, which $\Delta T_{zigzag(armchair)}$ is temperature drop in zigzag (armchair) part along the heat flux direction. Also, $G = \frac{J}{\Delta T_{in}}$ is the interface thermal conductance or the Kapitza conductance, which was calculated to be about $1.97\pm0.15 \, \frac{GW}{m^2 K}$ for the A-Z interface of A-Z GSNTs. In addition, $L=2L_{zigzag}(2L_{armchair})$ is the entire length of the A-Z GSNT.

**Results and discussion**

Single-wall silicene nanotubes were modeled by NEMD simulations with the purpose of finding the effect of length, diameter, temperature, grain boundary, random vacancy defects, and tensile strain on the phononic thermal conductivity.

Fig. 2 shows the temperature profiles of ASNT, ZSNT, and A-Z GSNT with the same length of 60 nm, which are plotted as the mean temperature of slabs perpendicular to the transport direction in the steady state condition. The mean temperature of each slab is calculated based on the following equation:

$$T_i = \frac{1}{3N_i k_B} \sum_{j=1}^{N_i} m_j \, v_j^2, \qquad (5)$$

where $T_i$ is the mean temperature of the slab i, $N_i$ is the number of Si atoms in the slab, $k_B$ is the Boltzmann constant and $m_j$ and $v_j$ are the mass and velocity of the jth atom in the slab, respectively. The temperature gradient is found by linear fitting to the temperature profile.

The result exhibits that the temperature distribution of nanotubes is linear in the area away from baths, but there is slight nonlinearity near the two ends because of phonon scattering with the heat baths. Also, a temperature gap with an approximate value of 3.8 K was observed in the temperature profile of A-Z GSNT. The discontinuity at the grain boundary is the signature of Kapitza resistance.

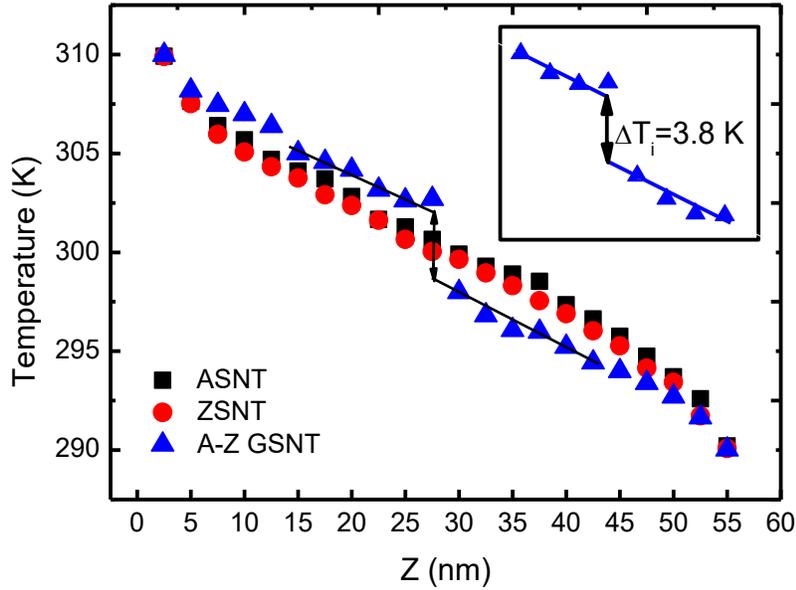

Fig. 2. The steady-state temperature profiles of ASNT, ZSNT, and A-Z GSNT with the same length of 60 nm and diameter of 15 nm at T=300 K and ΔT=20 K

Fig. 3 shows the accumulative energy added to and subtracted from the heat baths as a function of the time. The heat current $\frac{dE}{dt}$ is calculated as the slope of the linear fit to energy profile. It can be seen that the total energy of the system remains constant due to equality of the rate of added energy to the system and the rate of subtracted energy from that.

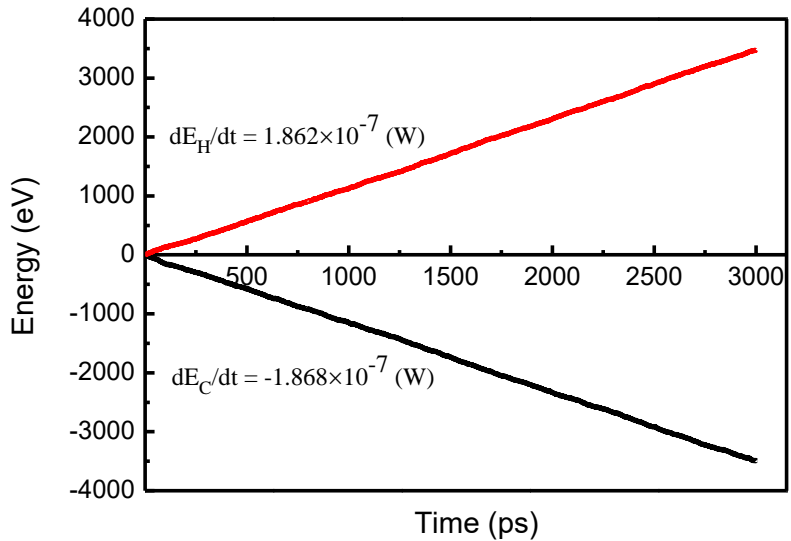

Fig. 3. Accumulative added energy to and subtracted energy from heat baths as a function of simulation time, to calculate the heat current belongs to ASNT with the diameter of 15 nm and the length of 60 nm at T=300 K and ΔT=20 K

We simulated silicene nanotubes in order to study the thermal conductivity as a function of length. The results are illustrated in Fig. 4 for nanotubes with length in the range of 40-150 nm and identical diameters of 15 nm. It is found that the thermal conductivity versus length of ASNTs and ZSNTs is obtained nearly in the range of 24-34 W/m.K, which is comparable with the thermal conductivity of single-layer silicene (40.1 W/m.K) and silicene nanoribbons (31.4-35.0 W/mK) [20]. These calculated values of the thermal conductivity, which are dramatically lower than that of carbon nanotubes (2000 W/m.K) and that of bulk silicon (165-238 W/m.K) [56,57], make silicene nanotubes a promising thermoelectric material.

Also, the results indicate that there is no significant difference between armchair and zigzag silicene nanotube in terms of thermal conductivity and silicene nanotubes almost show chirality independence in thermal conduction. Liu et al. [49] previously reported that the thermal conductivity of zigzag silicene nanosheet is just about 10% larger than that of armchair type; however, such phenomena are not observed in the case of silicene nanotubes.

According to Fig. 4, the thermal conductivities of silicene nanotubes possess a strong dependence on the length. With increasing the length up to 100 nm, the thermal conductivity increases remarkably and for the length of more than 100 nm the plot represents an approximate converging behavior. The explanation for this result is that the predicted thermal conductivity by NEMD method strongly depends on the system length along the heat flux direction. This size effect occurs because of the phonon scattering at the hot and cold baths and affects the heat transport in the system, which its length is smaller to the phonon mean-free path [58]. Thus, we investigate the effective phonon mean-free path of the silicene nanotubes by using the following equation [59]:

$$\frac{1}{\kappa} = \alpha \left(\frac{n}{L} + \frac{1}{l_\infty}\right) = \alpha \left(\frac{1}{l_{\text{eff}}}\right), \quad (6)$$

where $L$ is the length of silicene nanotube, $l_\infty$ is phonon mean-free path for the infinite system and $l_{\text{eff}}$ is effective phonon mean free path. Also, α is a function of group velocity and specific heat of phonons. The value of n is set to 2 for the systems of the nonperiodic condition. Equation (6) shows a linear relationship between the inverse of the thermal conductivity and the inverse of system length. Also, by extrapolation of this linear curve to $\frac{1}{L} = 0$, the thermal conductivity of an infinite system can be obtained.

The inset of Fig. 4 shows the inverse of thermal conductivity versus the inverse of system length for armchair and zigzag silicene nanotubes. A linear relationship between the $\frac{1}{\kappa}$ and $\frac{1}{L}$ can be observed. By linear extrapolating to $\frac{1}{L} = 0$, the thermal conductivity of an infinite armchair and zigzag silicene nanotubes can be obtained about 41±0.9 and 40±1.5 W/m.K, respectively. These values are in agreement with the thermal conductivity of silicene nanosheet measured by Pei et al. [21]. Also, the calculated Phonon mean-free path for an infinite armchair and zigzag nanotubes are 13.50 and 11.8 nm, respectively. It is worth noting that first-principles calculations have illustrated the largest phonon mean-free path of about 24 nm for silicene [60].

The effective phonon mean-free path is a parameter that is calculated for ASNTs and ZSNTs. Results are represented in Table 1. It is clear that effective phonon mean-free path increases with increasing the length of silicene nanotube and this increment is larger at smaller lengths.

Table 1. Effective phonon mean-free path for ASNTs and ZSNTs with different length

| Length(nm) | 40 | 50 | 60 | 70 | 80 | 90 | 100 | 110 | 120 | 130 | 140 | 150 |
|---|---|---|---|---|---|---|---|---|---|---|---|---|
| $l_{\text{eff}}$(AZNTs) | 8.05 | 8.76 | 9.31 | 9.74 | 10.9 | 10.38 | 10.62 | 10.83 | 11.02 | 11.17 | 11.31 | 11.44 |
| $l_{\text{eff}}$(ZSNTs) | 7.42 | 8.01 | 8.46 | 8.82 | 9.11 | 9.34 | 9.54 | 9.71 | 9.86 | 9.98 | 10.09 | 10.19 |

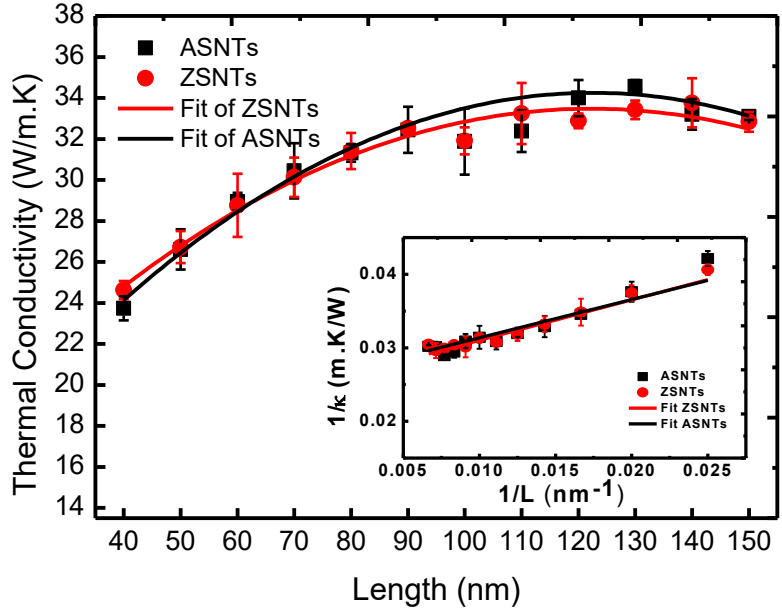

Fig. 4. Thermal conductivity versus length of ASNTs and ZSNTs; the lengths are within the range of 40-150 nm; diameters are set to 15 nm; and T=300 K and ΔT=20 K. The inset represents the extrapolation method used to obtain the thermal conductivity at $\frac{1}{L} = 0$.

Since the effect of diameter on the thermal conductivity of silicene nanotube received less attention, in this section, we evaluate the diameter dependence of the thermal conductivity in the armchair and zigzag type silicene nanotubes. Also, to evaluate the role of grain boundary on the thermal transport in silicene nanotubes, we applied a grain boundary between the armchair and zigzag oriented silicene nanotubes (A-Z GSNTs).

Fig. 5 presents the thermal conductivity versus diameter for nanotubes. Diameter changes from 8 nm to 15 nm and all structures have the same length equal to 100 nm. It is clear that the diameter range of 8-15 nm has no significant effect on the thermal conductivity of silicene nanotubes and by increasing the diameter, the thermal conductivity in all three types of silicene nanotubes changes slightly. These minor alternations are in the range of the statistical uncertainties of calculations. This finding is in good agreement with the previous report by Mortazavi et al. [61].

Furthermore, it was observed that although A-Z GSNTs exhibit similar diameter dependence thermal conductivity as ASNTs and ZSNTs, they show nearly 30% lower thermal conductivity compared with ASNTs and ZSNTs. The lower thermal conductivity is due to the existence of grain boundary and consequently phonon scattering; furthermore, this reduction in the thermal conductivity is because of temperature drop in the interface of the armchair-zigzag oriented nanotubes (Kapitza thermal resistance).

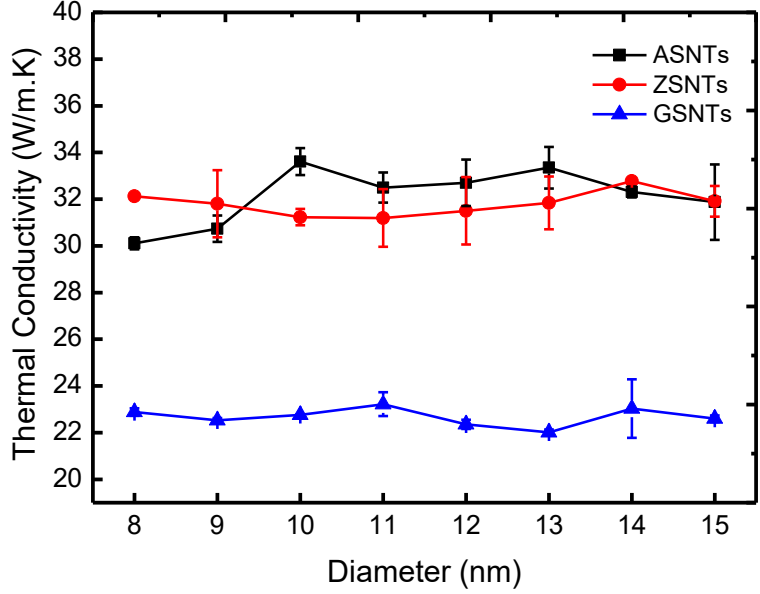

Fig. 5. Effect of diameter on the thermal conductivity of ASNTs, ZSNTs, and A-Z GSNTs with the same length of 100 nm and diameter in the range of 8-15 nm at T=300 K and ΔT=20 K

To have a better understanding of the impact of grain boundary on the thermal conductivity, the phonon power spectral density was investigated. The phonon power spectral density was determined by calculating the Fourier transform of the autocorrelation function of the velocity of atoms belonging to the two sides of the grain boundary (armchair and zigzag side) as follow [62-63]:

$$P(\omega) = \sum_i \frac{m_i}{k_B T_{MD}} \int_0^\infty e^{-i\omega t} <v_i(0).v_i(t)> dt, \qquad (7)$$

where $\omega$ is the angular frequency, $m_i$ is the mass of atom i and $v_i$ is the velocity of the ith atom.

Fig. 6 represents the phonon density of the two sides of A-Z GSNT (armchair and zigzag sides). It can be seen that there are mismatches between the two spectra of armchair side and zigzag side, which indicates phonon scattering in the grain boundary. Therefore, the reduction in the thermal conductivity of A-Z GSNTs compared with ASNTs and ZSNTs can be attributed to the existence of the grain boundary and consequently boundary thermal resistance.

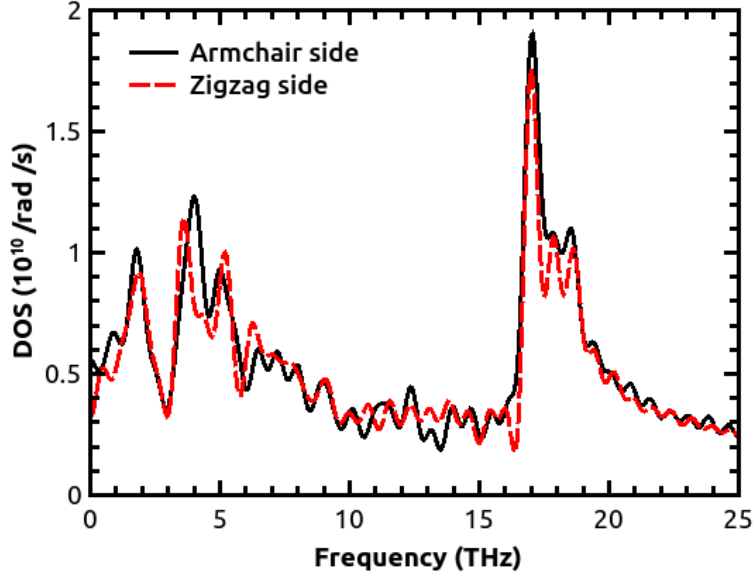

Fig. 6. Phonon power spectral density of the two sides of A-Z GSNT (armchair and zigzag sides)

To examine the sensibility of the thermal conductivity to the mean temperature of the silicene nanotubes, we enhance the mean temperature from 300 to 700 K. Two possibility occurs with increasing the mean temperature of a system; i.e., increasing the phonon-phonon scattering and increasing the excited high-energy (high frequency) phonons [41]. Fig. 7 presents that the thermal conductivity of the silicene nanotubes decreases by increasing the temperature. Thus, it is found that in these systems the competition between these mechanisms leads to suppressing excited phonons by phonon-phonon scattering.

Moreover, the inset in Fig. 7 represents that $T^{-1}$ dependence of the thermal conductivity for ASNT and ZSNT followed the phonon-phonon scattering. In fact, increasing temperature enhances the strength of phonon-phonon scattering, which is the dominant scattering mechanism especially in a perfect lattice. In the case of GSNT, due to the existence of grain boundary, both the phonon-phonon scattering and phonon-boundary scattering affect the thermal conductivity. Therefore, the curve deviates from the $T^{-1}$ law and A-Z GSNT exhibits lower thermal conductivity than that of the armchair and zigzag types.

Furthermore, it is clear that in lower temperature the sensibility of the thermal conductivity to the mean temperature is more than at a higher temperature. For instance, the thermal conductivity of ASNT and ZSNT decreases by about 50% when the temperature increases from 300-550 K while it decreases 18% in the temperature growth from 550-700 K.

We notice that the thermal conductivity of silicene nanotubes versus temperature is not sensitive to chirality thus the MD results obtained for ASNT and ZSNT are the same. Our findings are in agreement with previously research reported on the temperature dependence of the thermal conductivity of bulk silicene and silicene nanosheets [47, 49].

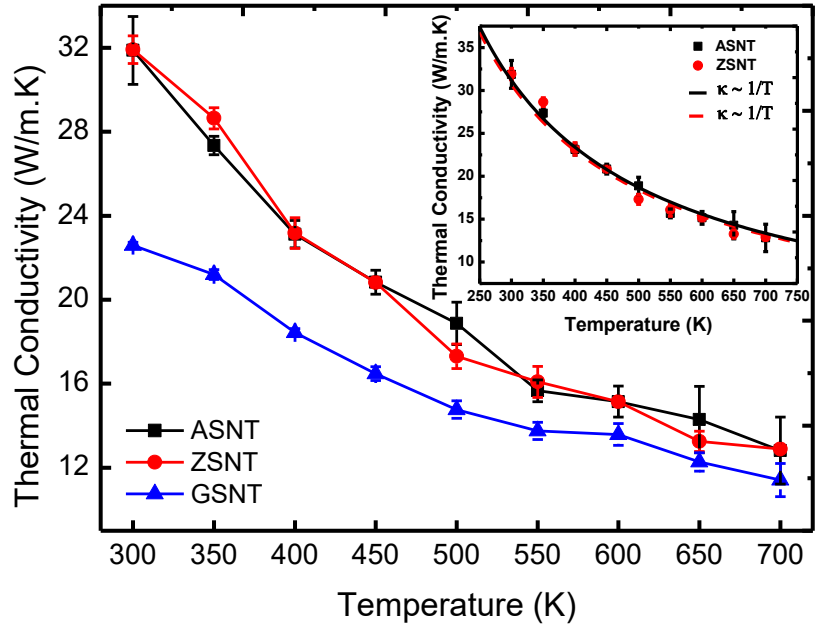

Fig. 7. Variation of the thermal conductivities of ASNT, ZSNT, and A-Z GSNT with temperature; all the tubes have the same length and diameter of 100 nm and 15 nm, respectively, and ΔT=20 K.

Studying the thermal properties of silicene nanotubes under extreme condition such as defects and strain is essential for practical applications. Thus, in this step, we examine the effect of vacancy defect on the thermal conductivity of silicene nanotubes because vacancy defect is inevitable during the fabrication process. We choose armchair and zigzag edge silicene nanotubes with the same size as the pristine structures and generate defects by random removal of the silicon atoms from the pristine structures.

The thermal conductivity of silicene nanotubes versus defect concentrations is represented in Fig. 8. It is clear that the thermal conductivity of defective silicene nanotubes exhibits a reduction with the increasing the defect concentration due to phonon-defect scattering. It is of note that the thermal conductivity as a function of vacancy defect is insensitive to the chirality of silicene nanotubes and the trend of plots for both armchair and zigzag types are the same.

Also, increasing the defect concentration up to nearly 2% reduces the thermal conductivity about 75%, which shows the pristine silicene nanotubes are highly sensitive to vacancy defect. As represented in Fig. 8, the plot of the thermal conductivity versus defect concentrations illustrates a converging behavior after approximately 2% of defect concentration. Our findings are in agreement with those reported by Li et al. for silicene nanosheets [64].

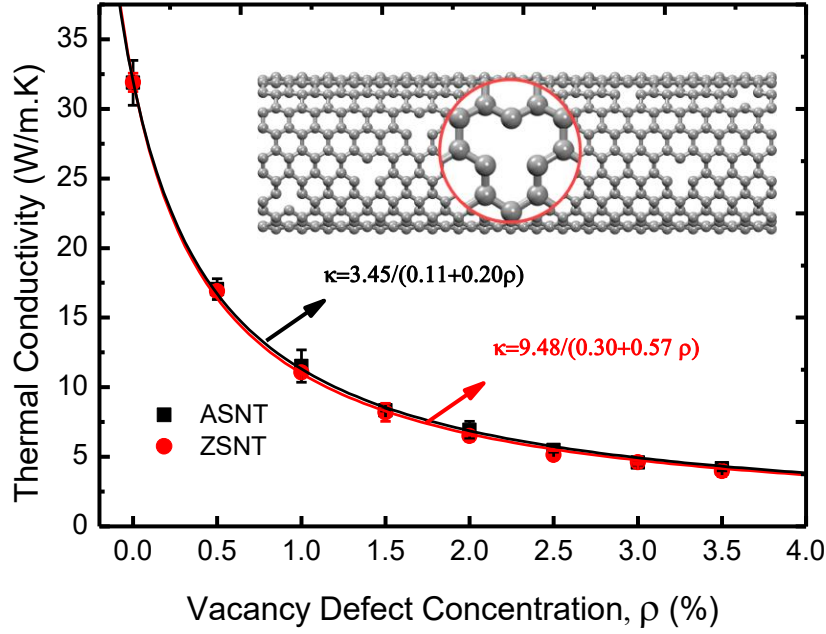

Fig.8. The thermal conductivity of armchair and zigzag silicene nanotubes versus vacancy defect at T=300 K and ΔT=20 K. The length and diameter of nanotubes are 100 nm and 15 nm, respectively.

In practice, nanostructures usually contain residual strain after fabrication that can impressively affect thermal properties of nanoscale devices [65, 66]. In the case of crystalline nanostructures, strain affects the stiffness of the nanostructures and consequently modifies their phonon group velocities [45]. Therefore, considering the importance of the impact of strain on the thermal properties of crystalline nanostructures, the effect of tensile strain from 0.01 to 0.1 on the thermal conductivity of ASNT, ZSNT, and A-Z GSNT was evaluated.
In this study, the strain is defined as:

$$\text{Strain} = \frac{dL}{L}, \qquad (8)$$

where $L$ is an initial length of the nanotube and $dL$ is the change in length due to stretching one or both ends of the nanotube along the longitudinal axis.
For applying the strain, the first slab of the nanotube is fixed, but the last slab starts to stretch along the nanotube length. The stretching velocity is set to 0.01 Å/ps. We schematically represent the applied strain on the ASNT in Fig. 9. As illustrated in this figure, the buckled structure of silicene starts to be smoothly flattened under tensile strain.

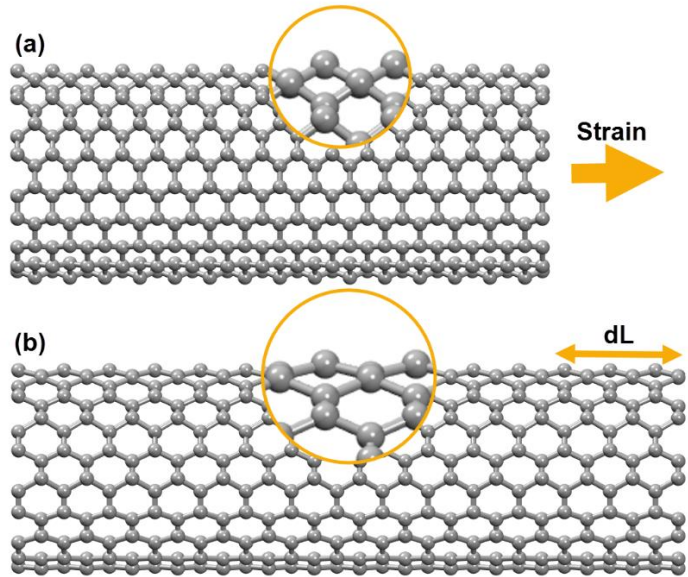

Fig. 9. (a) A schematic view of strain-free ASNT. (b) A schematic view of strained ASNT; the buckled structure of silicene starts to be smoothly flattened under tensile strain.

Fig. 10 shows the results of simulations for ASNT, ZSNT, and A-Z GSNT under the tensile strain. It is found that the thermal conductivity of the silicene nanotubes is sensitive to the strain. Also, the thermal conductivity of all three types of the nanotubes first increase with strain and then starts to decrease, which is comparable with previously reported values for silicene and silicene nanosheets [20, 21, 65]

These results are contrary to findings for graphene and carbon nanotubes. Generally, with increasing the strain, the thermal conductivity of graphene and carbon nanotubes decreases due to increasing the atomic bond length and consequently weakening the bond strength [66]. This distinct response of silicene nanotubes to tensile strain could be attributed to the buckled structure of silicene because by applying lower tensile strain, the bucked structure starts to become less buckled, leading to the increase in-plane stiffness of silicene and an enhanced thermal conductivity. It is worth noting that the first-principles calculation predicts silicene structure remains buckled under tensile strain even up to 12.5% [44]. Thus, by applying a significant amount of tensile strains, the buckled structure of silicene starts to become flattened and Si-Si bonds are stretched so that the in-plane stiffness of silicene and the thermal conductivity start to decrease [21].

In our study, the maximum thermal conductivity for ASNT and ZSNT is observed at strain values about 3% and 5%, respectively, which is about 28% more than that of the unstrained structure. For A-Z GSNT, it can be seen that the maximum thermal conductivity under 3% strain is about 27.5 W/m.K, which is 22% higher than strain-free A-Z GSNT. Surprisingly, thermal conductivity under 10% strain is nearly the same as the strain-free structure. As discussed before, the lower thermal conductivity of A-Z GSN compared with armchair and zigzag type nanotube is due to temperature drop in the grain boundary and interface thermal resistance.

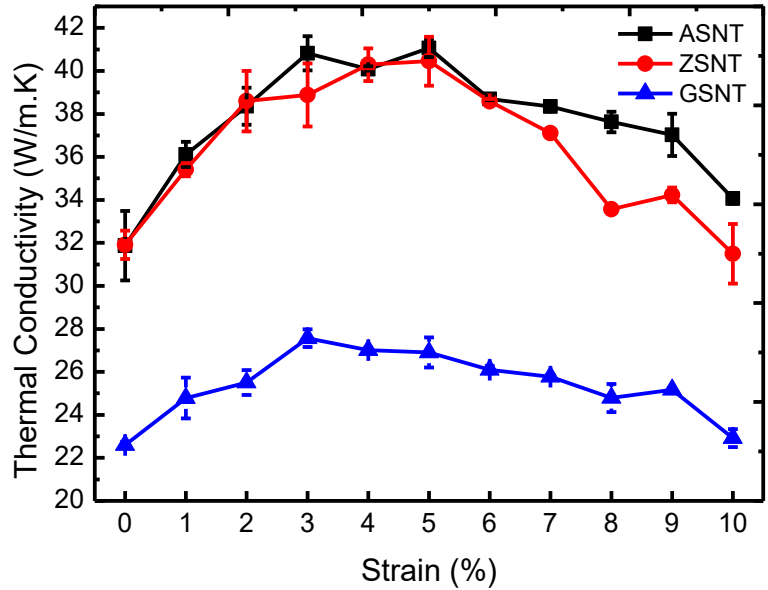

Fig. 10. The thermal conductivity of ASNT, ZSNT, and A-Z GSNT under tensile strain at T=300 K and ΔT=20 K; all the investigated nanotubes have the same length and diameter of 100 nm and 15 nm, respectively.

**Conclusions**

In this study, non-equilibrium molecular dynamics simulations were conducted to investigate the thermal conductivity of silicene nanotubes with different lengths, diameters, and chiralities by taking into account the effects of grain boundary, strain, vacancy defect, and temperature in the range of 300-700 K.

We found that the thermal conductivities of silicene nanotubes possess a strong dependence on the length such that its magnitude varies approximately in the range of 24-34 W/m.K, which is one order of magnitude lower than that of bulk silicon. Besides, our results have shown that the thermal conductivity of silicene nanotubes is independent of the chirality and diameter. We observed a linear relationship between the $1/\kappa$ and $1/L$ and obtained the thermal conductivity of the infinite armchair and zigzag silicene nanotubes as about 41 and 40 W/m.K, respectively. Furthermore, the calculated phonon mean-free paths for the infinite armchair and zigzag nanotubes were nearly 13.50 and 11.8 nm, respectively.

According to the result, the silicene nanotubes consisting of the grain boundary exhibit nearly 30% lower thermal conductivity compared with pristine ones due to the presence of grain boundary and consequently phonon scattering. The reduction in the thermal conductivity is due to the temperature drop in the interface (Kapitza thermal resistance). Furthermore, the thermal conductivity of armchair and zigzag silicene tubes decrease by about 60% as temperature rises from 300-700 K. In addition, with the increase in the vacancy defect concentration (due to phonon-defect scattering), we observed a significant reduction in the thermal conductivity of the system.

Finally, for strained nanotubes, we observed unexpected changes in the thermal conductivity so that the conductivity first increased significantly with tensile strain and then started to decrease. The maximum values of the thermal conductivity for the armchair and zigzag tubes were at the strains about 3% and 5%, respectively, which was about 28% more than that of the unstrained system. Also, for the system with grain boundary, the maximum thermal conductivity under 3% strain was about 27.5 W/m.K, which was 22% higher than strain-free structure. The different response of silicene nanotubes to tensile strain could be attributed to the buckled structure of the silicene.

*E-mail: khoeini@znu.ac.ir